\documentclass[prb,a4,
              twocolumn,
              showpacs,amsmath,amssymb]{revtex4-1}
\usepackage{epsfig}
\usepackage{graphicx}
\usepackage{subfigure}
\usepackage{color}
\usepackage{soul}
\usepackage{float}

\usepackage{epstopdf}
\usepackage{amsmath}
\usepackage{amssymb}
\usepackage{ifthen}
\usepackage{alltt}
\usepackage{fancyhdr}
\usepackage{verbatim}
\usepackage{moreverb}
\usepackage{rotating}

\usepackage{colortbl}
\usepackage{amssymb}
\usepackage{multirow}
\usepackage{bigstrut}
\usepackage{wrapfig}

\def\e{\begin{equation}}
\def\f{\end{equation}}
\def\_#1{{\bf #1}}

\def\.{\cdot}

\def\l#1{\label{eq:#1}}
\def\r#1{(\ref{eq:#1})}

\def\aeeo{\alpha_{\rm ee}^{\rm co}}
\def\aeer{\alpha_{\rm ee}^{\rm cr}}

\def\ammo{\alpha_{\rm mm}^{\rm co}}
\def\ammr{\alpha_{\rm mm}^{\rm cr}}

\def\=#1{\overline{\overline #1}}

\begin{document}

\title{One-way transparent sheets}
\author{Y.~Ra'di, V.~S.~Asadchy, and S.~A.~Tretyakov}
\affiliation{$^1$ Department of Radio Science and Engineering/SMARAD Center of Excellence, Aalto
University, P.O. Box 13000, FI-00076 AALTO, Finland}
\date{\today }

%%%%%%%%%%%%%%%%%%%%%%%%%%%%%%%%%%%%%%%%%%%%%%%%%%%%%%%%%%%%%%%%%%%%%%%%%%%%%%%%%%%%%%%%%%%%%%%%%%%%%%%%%%%%

\begin{abstract}
In this paper we introduce the concept of metasurfaces which are fully transparent when looking from one of the two sides of the sheet and have controllable functionalities for waves hitting the opposite side (one-way transparent sheets). We address the question on what functionalities are allowed, considering limitations due to reciprocity and passivity. In particular, we have found that it is possible to realize one-way transparent sheets which have the properties of a twist-polarizer in reflection or transmission when illuminated from the other side. Also one-way transparent sheets with controllable co-polarized reflection and transmission from the opposite side are feasible. We show that particular non-reciprocal magneto-electric coupling inside the sheet is necessary to realize lossless non-active transparent sheets. Furthermore, we derive the required polarizabilities of constituent dipole particles such that the layers composed of them form one-way transparent sheets. We conclude with design and simulations of an example of a nonreciprocal one-way transparent sheet functioning as an isolating twist-polarizer.
\end{abstract}

\maketitle

\section{Introduction}

%(called ``sheets'', if the thickness is electrically negligible)

Many novel and elegant device designs in antenna engineering and optics will become possible if we will be able to realize electrically (optically) thin layers  which have the application-required reflection and transmission coefficients. This need is addressed by inventing various metasurfaces (see  reviews in \cite{Holloway,metasurfaces}). Metasurfaces, usually realized as electrically and/or magnetically polarizable composite layers, can shape reflected and transmitted wavefront for required functionalities. Sheets with angle-stable
reflection and transmission \cite{Dienstfrey}, absorbing sheets \cite{Shvets,Watts,absorption}, high-impedance surfaces, including artificial magnetic conductors \cite{Sievenpiper}, and various polarization-transforming devices \cite{semchenko,euler10,alu,Teemu} are examples of metasurfaces. Recently, thin functional sheets have attracted considerable attention also in optics, since the possibilities to control optical transmission, reflection, and refraction using nonuniform sheets have been understood \cite{Gaburro,metasurfaces,Huygens1,Huygens2}. In majority of studies, the main focus has been on tailoring the transmitted wave while the reflection is kept as low as possible.

Here, we study sheets that are totally transparent from one of the two sides (\emph{one-way transparent sheets}) for normally incident plane waves. The main goal of this study is to find out what functionalities are possible if one-way transparency is required. Can we make the layer fully reflecting or act as a twist-polarizer or as a phase shifter for plane waves coming from the opposite, non-transparent side? The second related question is what kind of physical properties the constituents of these sheets must have in order to ensure the desired functionalities. Finally, we present a practical design example of a one-way transparent sheet. Clearly, one-way transparent sheets can have multiple applications due to their ``invisibility'' for excitations from one side.

Metasurfaces are microscopically structured layers (usually periodic), where the (average) distance between inclusions is smaller than the wavelength in the surrounding media, ensuring that the surfaces do not generate diffraction lobes. For an observer in the far zone the response is that of effectively homogeneous current sheets.
Thus, for layers of electrically negligible thickness (metasurfaces) illuminated by normally incident plane waves, the reflected and transmitted waves are plane waves created by the surface-averaged electric and magnetic current sheets with the surface current densities  $\_J_{\rm e}$ and $\_J_{\rm m}$. In composite sheets, the layer has a complicated microstructure, usually containing some electrically small but resonant inclusions (such as complex-shape patches or split rings or small helices). The surface-averaged current densities can be related to the electric and magnetic dipole moments $\_p$, $\_m$ induced in each unit cell as $\_J_{\rm e}={j\omega \_p\over S}$, $\_J_{\rm m}={j\omega \_m\over S}$. Here $S$ is the unit-cell area, and we use the time-harmonic convention $\exp(j\omega t)$. The higher-order multipoles induced in the inclusions do not contribute to the radiated plane-wave fields of the infinite array and we do not need to consider them explicitly. For realizing a one-way transparent sheet we will need to find such structures, where the induced surface-averaged  current densities equal zero for illumination from one of the sheet side but have non-trivial and controllable values if the incidence direction is reversed.
In the next section we will introduce the model for polarizations induced in the unit cells of metasurfaces which we will study here.

\section{Transparent arrays of bi-anisotropic unit cells}

\subsection{Effective polarizability dyadics of particles in periodic
arrays}

In order to reveal the most general possible functionalities of one-way transparent sheets, we assume the most general linear relations between the induced polarizations and the fields, the bi-anisotropic relations. It is convenient to write these relations as the linear relations between the dipole moments of the unit cells and the incident electromagnetic fields, which is equivalent to relating the induced surface current densities to the incident fields:
\begin{equation}
\left[ \begin{array}{c} \mathbf{p} \\ \mathbf{m}\end{array} \right]
=\left[ \begin{array}{cc}
\overline{\overline{\widehat{\alpha}}}_{\rm ee}&
\overline{\overline{\widehat{\alpha}}}_{\rm em}\\\overline{\overline{\widehat{\alpha}}}_{\rm me}&
\overline{\overline{\widehat{\alpha}}}_{\rm mm} \end{array} \right]\cdot
\left[ \begin{array}{c} \mathbf{E}_{\rm inc} \\ \mathbf{H}_{\rm
inc}\end{array} \right] . \label{eq:h}
\end{equation}
The effective polarizabilities include the
effects of lattice interactions and higher-order multipole excitations in unit cells.
In this paper, we consider isotropic (in the plane) thin sheets. The uniaxial symmetry ensures the isotropic response of the metasurfaces for normally-incident plane waves of arbitrary polarizations. The orientation of the layer in space is defined by the unit vector $\_z_0$, orthogonal to its plane.
The uniaxial symmetry allows only isotropic response and rotation
around the axis $\_z_0$. Thus, we can write all the polarizabilities in
(\ref{eq:h}) in the forms:
\begin{equation}
\begin{array}{c}\overline{\overline{\widehat{\alpha}}}_{\rm ee}=\widehat{\alpha}_{\rm ee}^{\rm co}\overline{\overline{I}}_{\rm t}+\widehat{\alpha}_{\rm ee}^{\rm cr}\overline{\overline{J}}_{\rm t},\qquad \displaystyle
\overline{\overline{\widehat{\alpha}}}_{\rm mm}=\widehat{\alpha}_{\rm mm}^{\rm co}\overline{\overline{I}}_{\rm t}+\widehat{\alpha}_{\rm mm}^{\rm cr}\overline{\overline{J}}_{\rm t},\vspace{.1cm}\\\displaystyle
\overline{\overline{\widehat{\alpha}}}_{\rm em}=\widehat{\alpha}_{\rm em}^{\rm co}\overline{\overline{I}}_{\rm t}+\widehat{\alpha}_{\rm em}^{\rm cr}\overline{\overline{J}}_{\rm t},\qquad\displaystyle
\overline{\overline{\widehat{\alpha}}}_{\rm me}=\widehat{\alpha}_{\rm me}^{\rm co}\overline{\overline{I}}_{\rm t}+\widehat{\alpha}_{\rm me}^{\rm cr}\overline{\overline{J}}_{\rm t},
\end{array}\label{eq:j}
\end{equation}
where indices ${\rm co}$ and ${\rm cr}$ refer to the symmetric and antisymmetric
parts of the corresponding dyadics, respectively.
Here, $\overline{\overline{I}}_{\rm t}=\=I-\_z_0\_z_0$ is the two-dimensional unit dyadic, and  $\overline{\overline{J}}_{\rm t}=\mathbf{z}_0\times\overline{\overline{I}}_{\rm t}$
is the vector-product operator.
In the last set of relations it is convenient to separate the coupling coefficients responsible for reciprocal and non-reciprocal coupling processes \cite{basic}:
\begin{equation}
\begin{array}{c}
\overline{\overline{\widehat{\alpha}}}_{\rm em}=(\widehat\chi-j\widehat\kappa)\overline{\overline{I}}_{\rm t}+(\widehat V+j\widehat \Omega)\overline{\overline{J}}_{\rm t},\vspace{.1cm}\\\displaystyle
\overline{\overline{\widehat{\alpha}}}_{\rm me}=(\widehat\chi+j\widehat \kappa)\overline{\overline{I}}_{\rm t}+(-\widehat V+j\widehat \Omega)\overline{\overline{J}}_{\rm t}.
\end{array}\label{eq:hof} \end{equation}
There are two
reciprocal classes (chiral $\widehat\kappa$ and omega $\widehat \Omega$) and two non-reciprocal classes (``moving'' $\widehat V$ and Tellegen $\widehat\chi$).
Note also that for reciprocal particles the electric and magnetic polarizabilities are always symmetric dyadics.  The four main types of magneto-electric coupling are summarized in Table~\ref{ta:main-classes}.

%%%%%%%%%%%%%%%%%%%%%%%%%%%%%%%%%%%%%%
%%%%%%%%%%%%%%%%%%%%%%%%%%%%%%%%%%%%%%
\begin{table}[h]
\caption{Magneto-electric coupling effects}
\begin{tabular}{|p{0.22\textwidth}|p{0.22\textwidth}|}
\hline
\rowcolor[gray]{.9}
\multicolumn{2}{|l|}{{\bf Types of magneto-electric coupling effects}}
\\
\hline
\rowcolor[gray]{.9}
%\centering
Omega
&
Chiral
\\
\hline
\vspace{0.1mm} $\begin{array}{c}
\displaystyle
\overline{\overline{\alpha}}_{\rm em}=\overline{\overline{\alpha}}_{\rm me}=j\Omega \overline{\overline{J}}_{\rm t}\\
%\vspace*{.5cm}\displaystyle
%\aemr\mbox{  is imaginary}
\end{array}$
&
\vspace{0.1mm} $\begin{array}{c}
 \displaystyle
\overline{\overline{\alpha}}_{\rm em}=-\overline{\overline{\alpha}}_{\rm me}=j\kappa\overline{\overline{I}}_{\rm t}\\
%\vspace*{.5cm}\displaystyle
%\aemo\mbox{  is imaginary}
\end{array}$
\vspace{.2cm}
 \\
\hline
\rowcolor[gray]{.9}
%\hspace{1.3cm}
%\centering
Moving
&
Tellegen
\\
\hline
\vspace{0.1mm}$\begin{array}{c}\displaystyle
\overline{\overline{\alpha}}_{\rm em}=-\overline{\overline{\alpha}}_{\rm me}=V\overline{\overline{J}}_{\rm t}\\
%\vspace*{.5cm}\displaystyle
%\aemr\mbox{is real}
\end{array}$
&
\vspace{0.1mm}$\begin{array}{c}\displaystyle
\overline{\overline{\alpha}}_{\rm em}=\overline{\overline{\alpha}}_{\rm me}=\chi\overline{\overline{I}}_{\rm t}\\
%\vspace*{.5cm}\displaystyle
%\aemo\mbox{  is real}
\end{array}$
\vspace{.2cm}
\\
\hline

\end{tabular}
\label{ta:main-classes}
\end{table}
%%%%%%%%%%%%%%%%%%%%%%%%%%%%%%%%%%%%%%
%%%%%%%%%%%%%%%%%%%%%%%%%%%%%%%%%%%%%%

The imaginary units in these notations are introduced in order to ensure that all the polarizability components are purely real for lossless particles. %This follows from the general conditions for lossless particles \cite{basic}:
%\begin{equation}
%\overline{\overline{\widehat\alpha}}_{\rm ee} = \overline{\overline{\widehat\alpha}}_{\rm ee}^\dag, \quad \overline{\overline{\widehat\alpha}}_{\rm mm} = \overline{\overline{\widehat\alpha}}_{\rm mm}^\dag, \quad \lt{and} \quad \overline{\overline{\widehat\alpha}}_{\rm em} = \overline{\overline{\widehat\alpha}}_{\rm me}^\dag ,
%\label{eq:t}
%\end{equation}
%where  $\dag$ is the Hermitian conjugate operator (complex conjugate and transposed matrix).
%

\subsection{Reflection and transmission of plane waves from uniaxial bi-anisotropic arrays}

We consider array properties for normally incident plane waves.
In the following theory of one-way transparent layers, we need to distinguish between
illuminations of the sheet from two opposite sides. In the rest of the paper, we
will use double signs for these two cases, where the top and bottom
signs correspond to the incident plane wave propagating in
$-\mathbf{z}_0$ and $\mathbf{z}_0$ directions, respectively. In the
incident plane wave, the electric and magnetic fields satisfy
\begin{equation}
\mathbf{H}_{\rm
inc}=\mp\frac{1}{\eta_0}\_z_0\times \mathbf{E}_{\rm
inc}=\mp\frac{1}{\eta_0}\overline{\overline{J}}_{\rm t}\.\mathbf{E}_{\rm
inc}, \label{eq:k}\end{equation}
where $\eta_0=\sqrt{\mu_0/\epsilon_0}$ is the wave impedance in the isotropic background medium (possibly free space).
In terms of the effective polarizabilities, the dipole moments in
(\ref{eq:h}) can be written as
\begin{equation}
\displaystyle \left[ \displaystyle\begin{array}{c} \mathbf{p} \\
\mathbf{m}\end{array} \right] =\left[\displaystyle
\begin{array}{c}\displaystyle
\overline{\overline{\widehat{\alpha}}}_{\rm ee}\mp\frac{1}{\eta_0}\overline{\overline{\widehat{\alpha}}}_{\rm em}\cdot\overline{\overline{J}}_{\rm t}\vspace{.1cm}\vspace*{.2cm}\\\displaystyle
\overline{\overline{\widehat{\alpha}}}_{\rm me}\mp\frac{1}{\eta_0}\overline{\overline{\widehat{\alpha}}}_{\rm mm}\cdot\overline{\overline{J}}_{\rm t}
\end{array}\right]\.\begin{array}{c}\mathbf{E}_{\rm inc}
\end{array}. \label{eq:l}\end{equation}
Knowing the dipole moments induced in each unit cell and substituting the polarizabilities from (\ref{eq:j}), we can now write the amplitudes of the reflected and transmitted plane waves as \cite{Teemu}

\begin{equation}
\begin{array}{l}
\displaystyle
\mathbf{E}_{\rm r}=-\frac{j\omega}{2S}[\eta_0\mathbf{p}\mp \mathbf{z}_0\times\mathbf{m}]
\vspace*{.2cm}\\\displaystyle
\hspace*{.5cm}=-\frac{j\omega}{2S}\left\{\left[\eta_0\widehat{\alpha}_{\rm ee}^{\rm co}\pm 2j\widehat\Omega -\frac{1}{\eta_0} \widehat{\alpha}_{\rm mm}^{\rm co}\right]\overline{\overline{I}}_{\rm t}\right.\vspace*{.2cm}\\\displaystyle
\hspace*{2cm}\left.+\left[\eta_0\widehat{\alpha}_{\rm ee}^{\rm cr}\mp 2\widehat\chi -\frac{1}{\eta_0} \widehat{\alpha}_{\rm mm}^{\rm cr}\right]\overline{\overline{J}}_{\rm t}\right\}\cdot\mathbf{E}_{\rm inc},
\end{array}
\label{eq:p}
\end{equation}

\begin{equation}
\begin{array}{l}
\displaystyle
\mathbf{E}_{\rm t}=\mathbf{E}_{\rm inc}-\frac{j\omega}{2S}[\eta_0\mathbf{p}\pm\mathbf{z}_0\times\mathbf{m}]\vspace*{.2cm}\\\displaystyle
\hspace*{.5cm}
=\left\{\left[1-\frac{j\omega}{2S}\left(\eta_0\widehat{\alpha}_{\rm ee}^{\rm co}\pm 2\widehat V
+\frac{1}{\eta_0}\widehat{\alpha}_{\rm mm}^{\rm co}\right)\right]\overline{\overline{I}}_{\rm t}\right.\vspace*{.2cm}\\\displaystyle
\hspace*{2cm}\left.
-\frac{j\omega}{2S} \left[\eta_0\widehat{\alpha}_{\rm ee}^{\rm cr}\mp 2j\widehat \kappa+\frac{1}{\eta_0} \widehat{\alpha}_{\rm mm}^{\rm cr}\right] \overline{\overline{J}}_{\rm t}\right\}\cdot\mathbf{E}_{\rm inc},
\end{array}\label{eq:q}
\end{equation}
in which, $S$ is the unit-cell area.
Using these relations, we will next study transparent layers.

\subsection{General conditions for totally transparent layers}
By definition, a transparent layer must not change the amplitude and phase of incident waves, that is,
\begin{equation}
\begin{array}{c}
\mathbf{E}_{\rm r}=0,\qquad\mathbf{E}_{\rm t}=\mathbf{E}_{\rm inc}.
\end{array}\label{eq:r}
\end{equation}
From (\ref{eq:p}) and (\ref{eq:q}) we find the necessary conditions for a transparent array of particles in the form
\begin{equation}
\begin{array}{c}
\displaystyle
\eta_0\widehat{\alpha}_{\rm ee}^{\rm co}\pm 2j\widehat\Omega -\frac{1}{\eta_0} \widehat{\alpha}_{\rm mm}^{\rm co}=0,
\vspace*{.2cm}\\\displaystyle
\eta_0\widehat{\alpha}_{\rm ee}^{\rm cr}\mp 2\widehat\chi -\frac{1}{\eta_0} \widehat{\alpha}_{\rm mm}^{\rm cr}=0,
\vspace*{.2cm}\\\displaystyle
\eta_0\widehat{\alpha}_{\rm ee}^{\rm co}\pm 2\widehat V
+\frac{1}{\eta_0}\widehat{\alpha}_{\rm mm}^{\rm co}=0,\vspace*{.2cm}\\\displaystyle
\eta_0\widehat{\alpha}_{\rm ee}^{\rm cr}\mp 2j\widehat \kappa+\frac{1}{\eta_0} \widehat{\alpha}_{\rm mm}^{\rm cr}=0.
\end{array}\label{eq:s}
\end{equation}
As above, the double signs correspond to the two opposite incidence directions. Clearly, these conditions ensure that the surface-averaged induced electric and magnetic current densities equal zero (see (\ref{eq:p}) and (\ref{eq:q})).

First, it is obvious that the conditions for two-way transparency allow only trivial solution: in this case all the polarizability components must equal zero. Indeed, if we demand that conditions (\ref{eq:s}) are satisfied for both choices of the $\pm $ sign, so that the sheet looks transparent from both sides, then all the magnetoelectric coefficients must be zero. Next, we see immediately that in that case all the other polarizabilities must also vanish. We stress that this does not imply that the sheet is simply absent: zero dipole moments in each unit cell mean only that the \emph{surface averaged} electric and magnetic current densities are zero. For example, a low-loss frequency-selective surface is transparent from both sides at the parallel resonance of the unit cell, although strong currents are induced in the structure. It is quite simple but interesting result. In particular, it implies that sheets which are fully transparent only from one side must exhibit electromagnetic coupling inside inclusions, or the sheet will be transparent from both sides. Note that this general conclusion holds also for non-reciprocal sheets.

What will be the properties of an array which is transparent from one side for waves coming from the other side? We expect that using different particles (reciprocal and non-reciprocal), we should be able to control the response seen from  the other side of the sheet.
Suppose, that a grid of particles is set to be transparent for $+\mathbf{z}_0$-directed incident waves. Then, using  (\ref{eq:s}),
we can express the electric and magnetic polarizabilities in terms of the magneto-electric parameters:
\begin{equation}
\begin{array}{c}
\displaystyle
\eta_0\widehat{\alpha}_{\rm ee}^{\rm co} -\frac{1}{\eta_0} \widehat{\alpha}_{\rm mm}^{\rm co}= 2j\widehat\Omega,
\vspace*{.2cm}\\\displaystyle
\eta_0\widehat{\alpha}_{\rm ee}^{\rm cr} -\frac{1}{\eta_0} \widehat{\alpha}_{\rm mm}^{\rm cr}=-2\widehat\chi,
\vspace*{.2cm}\\\displaystyle
\eta_0\widehat{\alpha}_{\rm ee}^{\rm co}
+\frac{1}{\eta_0}\widehat{\alpha}_{\rm mm}^{\rm co}= 2\widehat V,\vspace*{.2cm}\\\displaystyle
\eta_0\widehat{\alpha}_{\rm ee}^{\rm cr}+\frac{1}{\eta_0} \widehat{\alpha}_{\rm mm}^{\rm cr}=- 2j\widehat \kappa.
\end{array}\label{plus}
\end{equation}
From (\ref{eq:p}) and (\ref{eq:q}), the reflected and transmitted fields for the wave coming from the non-transparent side ($-\mathbf{z}_0$-directed wave) can be expressed in terms of the magneto-electric parameters only:
\begin{equation}
\begin{array}{l}
\displaystyle
\mathbf{E}_{\rm r}=\frac{j2\omega}{S}\left(-j\widehat\Omega\overline{\overline{I}}_{\rm t} +\widehat \chi \overline{\overline{J}}_{\rm t}\right)\cdot\mathbf{E}_{\rm inc},
\end{array}
\label{eq:u}
\end{equation}
\begin{equation}
\begin{array}{l}
\displaystyle
\mathbf{E}_{\rm t}=\left[ \left(1 - \frac{j2\omega}{S}\widehat V\right)\overline{\overline{I}}_{\rm t}-\frac{2\omega}{S} \widehat \kappa\overline{\overline{J}}_{\rm t}\right]\cdot\mathbf{E}_{\rm inc}.
\end{array}\label{eq:v}
\end{equation}
Now, we can study possible responses of reciprocal and non-reciprocal one-way transparent sheets.

\subsection{Reciprocal one-way transparent sheets}

Let us first consider arrays of reciprocal unit cells (omega and chiral bi-anisotropic coupling).
For reciprocal particles, the electric and magnetic polarizabilities are symmetric dyadics ($\widehat \alpha_{\rm ee}^{\rm cr}=0$, $\widehat \alpha_{\rm mm}^{\rm cr}=0$), and the parameters of the non-reciprocal magneto-electric coupling vanish ($\widehat \chi=\widehat V=0$). The last relation in (\ref{plus}) tells that the chirality parameter  $\widehat\kappa$ is also zero.
This ensures that the transmission coefficient from the other side (\ref{eq:v}) equals unity, as it should be due to reciprocity.

From the other relations (\ref{plus}) we find that
\begin{equation}
\begin{array}{c}
\displaystyle
\eta_0\widehat{\alpha}_{\rm ee}^{\rm co}=j\widehat \Omega=-\frac{1}{\eta_0} \widehat{\alpha}_{\rm mm}^{\rm co}
\end{array}\label{eq:w}
\end{equation}
and, using (\ref{eq:u}), the reflected  field for the waves coming from the other side can be written as
\begin{equation}
\begin{array}{l}
\displaystyle
\mathbf{E}_{\rm r}=\frac{2\omega}{S}\widehat \Omega \, \mathbf{E}_{\rm inc}.
\end{array}
\label{eq:x}
\end{equation}
If the array is passive, then the absolute value of this reflection coefficient must equal zero, because the transmission coefficient is unity from both sides (due to reciprocity). Thus, the omega coupling coefficient in passive reciprocal one-way transparent sheets must be zero for all non-zero frequencies, and we end up with the trivial solution when all the polarizabilities are zero. However, if the inclusions can be active, we see that reciprocal layers can be transparent from one side, while the co-polarized reflection from the other side can be controlled by the value of the omega coupling. Note that this is the only possible functionality even for active inclusions: The requirement of reciprocity is very limiting, because it sets the transmission coefficient to be unity from both sides. In particular, this does not allow chirality in the particles, and, thus, no polarization transformation is possible in isotropic reciprocal one-way transparent sheets.

\subsection{Non-reciprocal one-way transparent sheets}

As is evident from equations (\ref{eq:u}) and (\ref{eq:v}), the use of non-reciprocal particles in principle allows full control over co- and cross-polarized reflection and transmission coefficients of one-way transparent sheets (passivity limitations are discussed below). To find out what polarizabilities are required for any desired functionality one can start from  the required reflection and transmission coefficients and find the corresponding magneto-electric parameters. For example, if we would like to realize a one-way transparent twist-polarizer in transmission ($\_E_{\rm t}=\_z_0\times \_E_{\rm inc}=\=J_{\rm t}\cdot \_E_{\rm inc}$), the required values of the coupling parameters read
\begin{equation}
\widehat\kappa=-{S\over 2\omega},\qquad \widehat V=-j{S\over 2\omega},\qquad \widehat \Omega=\widehat\chi=0.
\end{equation}
The corresponding electric and magnetic polarizabilities follow from  (\ref{plus}):
\begin{equation}
\begin{array}{c}
\eta_0\widehat{\alpha}_{\rm ee}^{\rm co}=\frac{1}{\eta_0} \widehat{\alpha}_{\rm mm}^{\rm co}=\widehat V=-j{S\over 2\omega},\vspace{.2cm}\\\displaystyle
\eta_0\widehat{\alpha}_{\rm ee}^{\rm cr}=\frac{1}{\eta_0} \widehat{\alpha}_{\rm mm}^{\rm cr}=-j \widehat\kappa =j{S\over 2\omega}.
\end{array}\label{eq:one1}
\end{equation}

Note that the magnitudes of all the required normalized polarizabilities are equal, and this provides one of the examples of extreme response of balanced bi-anisotropic particles \cite{balanced}. It is easy to check from (\ref{eq:l}) that in this case of zero reflection the induced electric and magnetic dipoles of each unit cell form Huygens pairs, radiating only in the forward direction, as it should be for any non-reflecting sheet (see examples in \cite{Teemu,Huygens1,Huygens2}).

As a second example, let us consider the one-way transparent twist-polarizer in reflection. Now we set $\widehat \Omega=0$, so that the reflected field is twist-polarized. Choosing the value of the Tellegen parameter to be $\widehat \chi=-j{S\over 2\omega}$, the amplitude and phase of the reflected cross-polarized field are equal to those of the incident field. The first two equations in (\ref{plus}) tell that the normalized co-polarized electric and magnetic polarizabilities are equal and that we should have at least one of the antisymmetric components in electric and magnetic polarizabilities non-zero.
If the device is passive, the amplitude of the transmitted field must be zero (because the reflected field has the same amplitude as the incident field). This determines the chirality parameter $\widehat \kappa=0$ and the velocity parameter $\widehat V=-j{S\over 2\omega}$. This gives a unique solution for the electric and magnetic polarizabilities:
\begin{equation}
\begin{array}{c}
\eta_0\widehat{\alpha}_{\rm ee}^{\rm co}= \frac{1}{\eta_0} \widehat{\alpha}_{\rm mm}^{\rm co}=\widehat V=-j{S\over 2 \omega},
\vspace{.5cm}\\\displaystyle
\eta_0\widehat{\alpha}_{\rm ee}^{\rm cr}=-\frac{1}{\eta_0} \widehat{\alpha}_{\rm mm}^{\rm cr}=-\widehat \chi=j{S\over 2 \omega}.
\end{array}\end{equation}
It is interesting to notice that also in this case the values of all normalized parameters are equal (another example of  balanced bi-anisotropic particles).

As is already clear from the above, the requirement of passivity of the particles imposes certain limitations on achievable response. For example, if we assume that only the Tellegen parameter $\widehat \chi$ is non-zero while all the other coupling coefficients are zero, we see that the transmission coefficient from both sides equals unity, and conclude that in this case it is possible to control the cross-polarized reflection only if the particles are active.

An interesting case is the case of a ``moving'' grid ($\widehat{V}\neq 0$). We can set $\widehat \Omega=\widehat \chi=0$, so that the reflection coefficient is zero and the induced current sheets form a Huygens' pair. Upon substitution of $\widehat \Omega=\widehat \chi=0$ in (\ref{plus}), we see that the electric and magnetic polarizabilities are balanced:
\begin{equation}
\begin{array}{c}
\displaystyle
\eta_0\widehat{\alpha}_{\rm ee}^{\rm co}= \frac{1}{\eta_0} \widehat{\alpha}_{\rm mm}^{\rm co},\vspace{.2cm}\\\displaystyle
\eta_0\widehat{\alpha}_{\rm ee}^{\rm cr}= \frac{1}{\eta_0} \widehat{\alpha}_{\rm mm}^{\rm cr}.
\end{array}\l{movco}
\end{equation}

 We conclude that we can fully control the transmission coefficient choosing the values of $\widehat V$ and $\widehat \kappa$ (each of these two parameters will uniquely define the values in \r{movco}), maintaining the property of zero reflection (Huygens' layer).
The only limitation on the transmission coefficient values comes from passivity: The total amplitude of the transmitted field should be smaller than the amplitude of the incident field. One of the interesting limiting cases is the case of non-chiral ``moving'' arrays. Setting $\widehat \kappa=0$, we find the required effective polarizabilities as
\begin{equation}
\begin{array}{c}
\displaystyle
\eta_0\widehat{\alpha}_{\rm ee}^{\rm co}= \widehat V=\frac{1}{\eta_0} \widehat{\alpha}_{\rm mm}^{\rm co},\vspace*{.2cm}\\\displaystyle
\eta_0\widehat{\alpha}_{\rm ee}^{\rm cr}=\frac{1}{\eta_0} \widehat{\alpha}_{\rm mm}^{\rm cr}=0
\end{array}
\label{eq:aa}
\end{equation}
and, using (\ref{eq:u}) and (\ref{eq:v}), the reflection and transmission coefficients for the wave coming from the other side can be written as
\begin{equation}
\begin{array}{l}
\displaystyle
\mathbf{E}_{\rm r}=0,\qquad\displaystyle
\mathbf{E}_{\rm t}=\left(1 -\frac{j2\omega}{S}\widehat V\right){\_E}_{\rm inc}.
\end{array}
\label{eq:bb}
\end{equation}
One can see that the sheet of ``moving'' particles can be designed to work as
a completely transparent layer from one side and a partially transparent layer from the other side (with controllable amplitude and phase of the transmitted field).
In the special case of a balanced and lossy layer the sheet is transparent from one side and acts as a perfect absorber from the other side \cite{absorption}.

\section{Requirements for individual polarizabilities of unit cells}

The above theory gives the required conditions for effective (collective) polarizabilities of unit cells forming one-way transparent sheets. These parameters connect the induced electric and magnetic dipole moments to the incident electric and magnetic fields, see (\ref{eq:h}). Thus, the effective polarizabilities depend not only on the individual particles but also on electromagnetic coupling between particles in the infinite array. Here, we use the known theory of reflection and transmission in infinite dipole arrays (e.g., \cite{analytical}) to find the corresponding requirements on the polarizabilities of individual particles in free space. This is necessary to approach the problem of the particle design (finding the inclusion shape and sizes which provide the desired response of the whole array).
To characterize individual particles, we consider their response to the  local
electromagnetic fields, which exist at the position of one reference particle:
\begin{equation}
\left[ \begin{array}{c} \mathbf{p} \\ \mathbf{m}\end{array} \right]
=\left[ \begin{array}{cc} \overline{\overline{\alpha}}_{\rm ee}& \overline{\overline{\alpha}}_{\rm em}\\\overline{\overline{\alpha}}_{\rm me}& \overline{\overline{\alpha}}_{\rm mm} \end{array} \right]\.\left[ \begin{array}{c} \mathbf{E}_{\rm loc} \\ \mathbf{H}_{\rm loc}\end{array} \right].
\label{eq:e}\end{equation}
Since the grid
is excited by plane-wave fields which are uniform in the array plane (normal
incidence), the induced dipole moments are the
same for all particles.
The local fields exciting the particles are the sums of the external
incident fields and the interaction fields caused by the induced
dipole moments in all other particles:
\begin{equation}
\begin{array}{c}
\mathbf{E}_{\rm loc}=\mathbf{E}_{\rm inc}+\beta_{\rm e}\, \mathbf{p},
\vspace*{.2cm}\\\displaystyle
\mathbf{H}_{\rm loc}=\mathbf{H}_{\rm inc}+\beta_{\rm m}\,\mathbf{m},
\end{array}\label{eq:f}
\end{equation}
 where $\beta_{\rm e}$ and $\beta_{\rm m}$ are the interaction constants that describe the effect of the entire array on a single inclusion. These dyadic coefficients are proportional to the two-dimensional unit dyadic $\={I}_{\rm t}$. Explicit analytical expression for the interaction constants can be found in \cite{basic}. 

Because all of the dipoles are in the same plane, the induced magnetic dipoles do not produce any electric interaction field in the tangential plane, and vice versa (see \cite{yatsenko03}).
Expressing the incident fields in (\ref{eq:h}) in terms of the local fields and the interaction constants (\ref{eq:f}) we can find find the polarizabilities of the individual unit cells in terms of the required collective polarizabilities. In the general case of uniaxial polarizabilities these expressions can be found in \cite{Teemu}.
As an example, let us study the case of one-way transparent non-chiral ``moving'' sheet (the required effective polarizabilities are given by (\ref{eq:aa})).  In this case the relations between the individual and collective polarizabilities can be written as
\begin{equation}
\begin{array}{c}
\displaystyle
\overline{\overline{\widehat{\alpha}}}_{\rm ee}=\frac{\aeeo-\beta_{\rm m}(\aeeo\ammo-V^2)}{1-(\aeeo\beta_{\rm e}+\ammo\beta_{\rm m})+\beta_{\rm e}\beta_{\rm m}(\aeeo\ammo-V^2)}\=I_{\rm t},
\vspace*{.2cm}\\\displaystyle
\overline{\overline{\widehat{\alpha}}}_{\rm mm}=\frac{\ammo-\beta_{\rm e}(\aeeo\ammo-V^2)}{1-(\aeeo\beta_{\rm e}+\ammo\beta_{\rm m})+\beta_{\rm e}\beta_{\rm m}(\aeeo\ammo-V^2)}\=I_{\rm t},
\vspace*{.2cm}\\\displaystyle
\widehat{V} \hspace{1mm} \overline{\overline{J}}_{\rm t}=\frac{V}
{1-(\aeeo\beta_{\rm e}+\ammo\beta_{\rm m})+\beta_{\rm e}\beta_{\rm m}(\aeeo\ammo-V^2)}\overline{\overline{J}}_{\rm t}.
\end{array}\label{eq:ff}
\end{equation}
The necessary conditions for the polarizabilities of the single moving particle can be found substituting (\ref{eq:ff}) in conditions (\ref{eq:aa}) as
\begin{equation}
\begin{array}{c}
\displaystyle
\eta_0\aeeo=V=\frac{1}{\eta_0} \ammo,
\vspace*{.2cm}\\\displaystyle
\aeer= \ammr=0.
\end{array}
\label{eq:gg}
\end{equation}
These conditions are the balance conditions for individual particles. Interestingly, conditions (\ref{eq:gg}) were obtained in paper \cite{Joni} as the conditions for zero total scattering from small single uniaxial nonreciprocal particles.

Finally, we note that the required reciprocal magneto-electric coupling in particles can be realized using proper shaping of metal or dielectric inclusions (helical shapes for chiral particles and, for example, the shape of the letter $\Omega$ for omega coupling). Non-reciprocal coupling requires the use of non-reciprocal components, such as magnetized ferrite or plasma, or active components, e.g., amplifiers. For details we refer to  \cite{basic,lindell94} and references therein.

\section{Example of a one-way transparent sheet composed of non-reciprocal bi-anisotropic particles}

\begin{figure}[H]
\centering
\epsfig{file=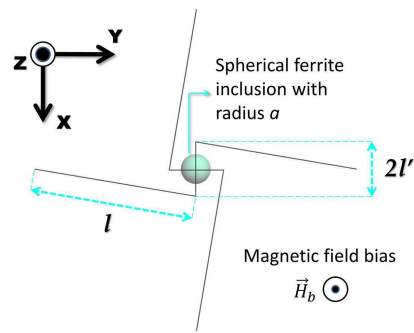, width=0.3\textwidth}
\caption{Geometry of a ``chiral-moving" particle. The external magnetic field bias is along the $\mathbf{z}_0$-axis.}
\label{ris:particle}
\end{figure}

\begin{figure}[H]
\centering
\includegraphics[width=\columnwidth]{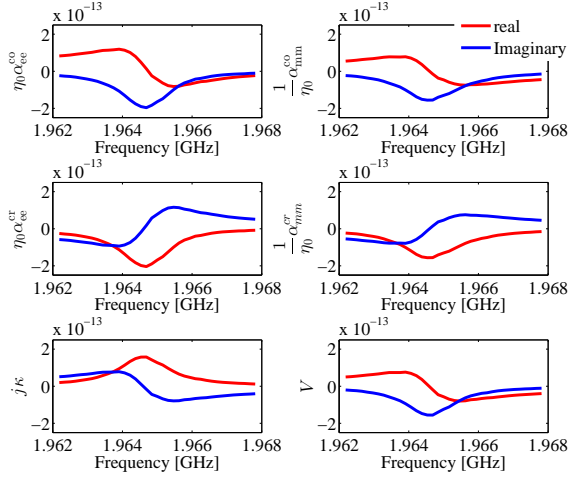}
\caption{Simulated polarizabilities of individual particle.}
\label{fig:individual}
\end{figure}

\begin{figure}[H]
\centering
\includegraphics[width=\columnwidth]{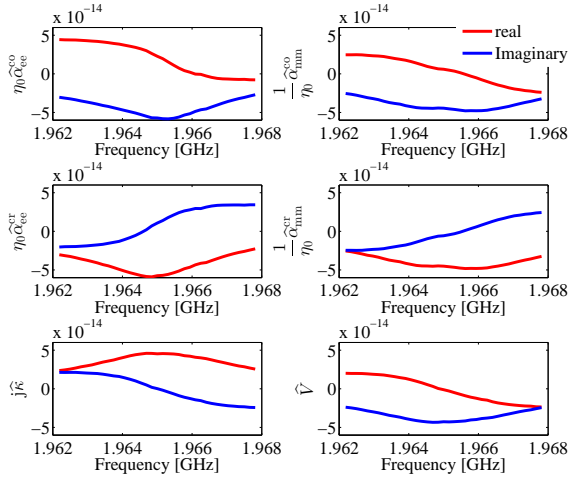}
\caption{Simulated effective polarizabilities of the grid.}
\label{fig:effective}
\end{figure}

\begin{figure}[H]
\centering
\includegraphics[width=0.95\columnwidth]{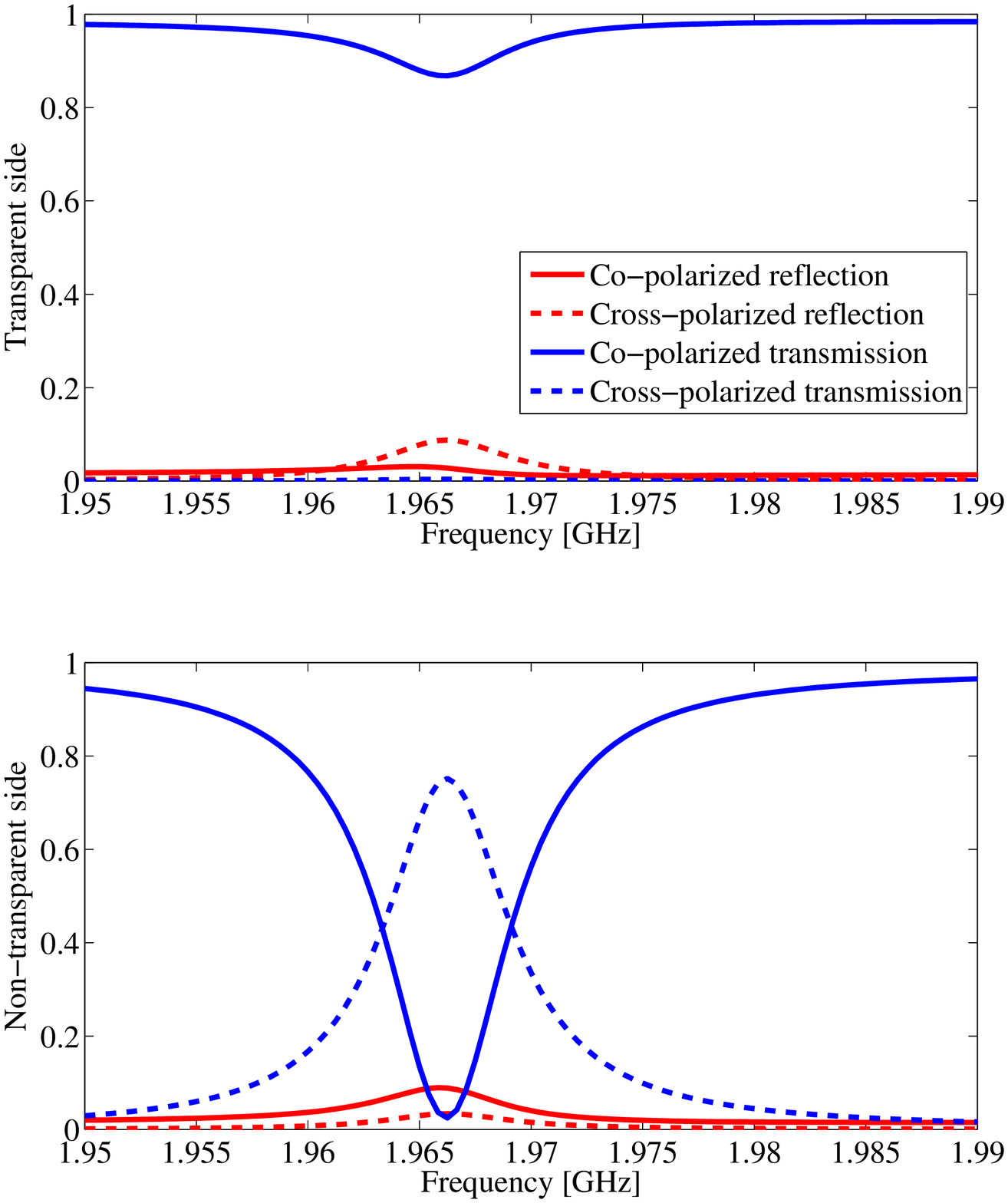}
\caption{Simulated reflection and transmission (in terms of intensity) for the sheet when the incident wave propagates along (a) $+\mathbf{z}_0$-axis and (b) $-\mathbf{z}_0$-axis.}%
\label{fig:RT}
\end{figure}

In this section, we present a realizable design of a one-way transparent sheet acting as a twist-polarizer from the non-transparent side. We verify the operation of the sheet by full-wave simulations using the Ansoft High Frequency Structure Simulator (HFSS). 
%As discussed above, it is impossible to design a one-way transparent sheet made of non-active reciprocal inclusions. Nevertheless, there is an opportunity to design such a sheet using passive  non-reciprocal electrically and magnetically polarizable particles. 
As a non-reciprocal particle, we use the particle possessing ``chiral-moving" coupling (bi-anisotropy parameters $\widehat \Omega=\widehat \chi=0$) presented in \cite{particle}. As it is shown in the Fig.~\ref{ris:particle}, the particle includes a ferrite sphere magnetized by external bias field and coupled to metal elements. Recently, polarizabilities of this particle were extracted analytically and numerically \cite{Sajad}.
As it was noted above, for the one-way transparency operation, the sheet must have balanced effective polarizabilities. We use the numerical method introduced in \cite{Fanyaev} which allows us to extract polarizabilities of an arbitrary polarizable particle. Using this method, parameters of the particle are optimized. The optimized dimensions of the particle (the target frequency is about $2$ GHz) read:
\mbox{$l=18$ mm}, \mbox{$l'=3$ mm}, \mbox{$a=1.65$ mm}, and the radius of the wire is \mbox{$r_0=0.05$ mm}. Material of the metal elements is copper and the ferrite material is yttrium iron garnet.
The properties of the ferrite material are:
the relative permittivity $\epsilon_r=\, 15$, the dielectric loss tangent $\tan\delta= 10^{-4}$, saturation magnetization $M_S=1780$ G and the full resonance linewidth $\Delta H=0.2$ Oe (measured at $9.4$ GHz). The internal bias field is $H_b=9626$ A/m, corresponding to the desired resonance frequency.
Simulated individual polarizabilities of the optimized particle and effective polarizabilities of the grid (area cell $S=1482.25$ mm$^2$) are shown in Figs.~\ref{fig:individual} and \ref{fig:effective}, respectively.
These figures exhibit fairly balanced electric and magnetic response in terms of individual and effective polarizabilities. Also it can be seen that at the resonance frequency the effective polarizabilities do not satisfy the conditions (\ref{eq:one1}) but they satisfy the following similar conditions:
\begin{equation}
\eta_0\widehat{\alpha}_{\rm ee}^{\rm co}=\frac{1}{\eta_0} \widehat{\alpha}_{\rm mm}^{\rm co}=\widehat V=-j{S\over 2\omega},
\label{eq:one11}
\end{equation}
\begin{equation}
\eta_0\widehat{\alpha}_{\rm ee}^{\rm cr}=\frac{1}{\eta_0} \widehat{\alpha}_{\rm mm}^{\rm cr}=-j \widehat\kappa =-{S\over 2\omega}.
\label{eq:two22}
\end{equation}
Conditions (\ref{eq:one11}) and (\ref{eq:two22}) are related to the case of a one-way transparent sheet which acts as twist-polarizer with additional $90^\circ$ phase shift for the non-transparent side ($\_E_{\rm t}=j \=J_{\rm t}\cdot \_E_{\rm inc}$). To calculate reflection and transmission of the sheet, periodic boundary conditions. Simulated co- and cross-polarized reflection and transmission for the wave incident from the transparent and non-transparent sides are shown in Fig.~\ref{fig:RT}.

As is seen from Fig.~\ref{fig:RT}, the designed composite sheet transmits 87\% of the incident wave power propagating along the transparent direction. At the same time, it transmits 75\% of the wave power (in cross polarization) when the layer is illuminated from the non-transparent side. Thus, the sheet acts as a one-way transparent layer and as twist-polarizer from the non-transparent side, as it was predicted theoretically. Non-ideal magnitude of the cross-polarized transmitted wave can be explained by inevitable absorption loss inside the ferrite sphere and copper wires (about 13\%). Also, some small reflection exists due to parasitic weak omega and Tellegen coupling effects in the particle (about 12\%).

The designed sheet is a non-reciprocal analogy of the device proposed in \cite{Teemu} which consists of reciprocal chiral particles and acts as a twist-polarizer for both directions of incidence. Non-reciprocal electromagnetic coupling allows us to obtain dramatically different response for the opposite incident directions.
The proposed composite sheet based on  ``chiral-moving" particles exhibits the target electromagnetic properties of a one-way transparent sheet and has realistic parameters allowing practical realizations.

%The other proposed in this paper non-active one-way transparent sheets (pure "moving" and "Tellegen-moving") can be designed basing on other non-reciprocal particles whose topologies are in prospect to find.

\section{Conclusion}
Although it is not possible to realize a fully transparent sheet except the trivial case of zero averaged induced surface currents, we have shown that it is possible to realize one-way transparent sheets. In these structures,  the polarizabilities of unit cells are different from zero, but they are balanced in such a way, that the averaged induced currents are zero for illumination from one of the two sides of the sheet. However, the response to plane waves illuminating the opposite side of the sheet is non-trivial and can be controlled by design of the metasurface microstructure.
Electromagnetic coupling (bi-anisotropy) inside unit cells of the metasurface is a necessary condition for one-way transparent layers.
If we are limited to the case of lossless sheets with passive particles, then one-way transparency necessarily requires non-reciprocal coupling and is impossible with chiral and omega particles. It was shown that presence of ``moving'' coupling is necessary to make a lossless non-active one-way transparent sheet. In particular, it has been shown that non-reciprocal coupling effects allow to realize a one-way transparent sheet which acts as a twist-polarizer in reflection or transmission when illuminated from non-transparent side. Another possible device is a one-way transparent phase-shifting sheet. If active particles are allowed, our possibilities to control electromagnetic response from the opposite side of the sheet are extended: There is no restriction on the amplitude of reflection and transmission for the wave coming from the non-transparent side. Also omega coupling becomes allowed and makes it possible to realize a one-way transparent sheet with controllable co-polarized reflection from the opposite side.
Required effective and individual polarizabilities of bi-anisotropic particles as components of a one-way transparent layer have been derived. Finally, we have shown a realistic design of a non-reciprocal one-way transparent sheet and simulated its performance parameters.

%\section*{Acknowledgements}

\end{document}